\documentclass[a4paper,preprint,12pt]{elsarticle}

\makeatletter
\def\ps@pprintTitle{%
  \let\@oddhead\@empty
  \let\@evenhead\@empty
  \let\@oddfoot\@empty
  \let\@evenfoot\@oddfoot
}
\makeatother

\usepackage[utf8]{inputenc}
\usepackage{natbib}
\usepackage{graphicx}
\usepackage{float}
\usepackage{siunitx}
\usepackage[english]{babel}
\usepackage[usenames,dvipsnames]{color}
\usepackage[none]{hyphenat}
\usepackage{verbatim}
\usepackage{setspace}
\usepackage{lineno}
\usepackage{adjustbox}
\usepackage{amssymb,amsmath,amsthm} 
\usepackage{enumerate}  
\usepackage{enumitem} 
\usepackage{subfig}

\renewcommand{\Im}{\hspace{2pt}\text{Im}}
\renewcommand{\Re}{\hspace{2pt}\text{Re}}
\newcommand{\sgn}{\text{sgn}}
\renewcommand{\sin}{\hspace{2pt}\text{sin}}

\begin{document}

\begin{frontmatter}

\title{Heat transport and cooling performance in a nanomechanical system with local and non local interactions.}

\author[1]{N. Beraha\corref{cor1}}
\cortext[cor1]{First corresponding author}
\ead{nberaha@campus.ungs.edu.ar}
\author[2,3]{A. Soba}
\author[1,3]{M. F. Carusela}

\address[1]{Instituto de Ciencias, Universidad Nacional de Gral. Sarmiento, Los Polvorines, Buenos Aires, Argentina}
\address[2]{Centro At\'omico Constituyentes - CNEA, Buenos Aires, Argentina}
\address[3]{Consejo Nacional de Investigaciones Cient\'\i ficas y T\'ecnicas, Argentina}

\begin{abstract}
In the present work, we study heat transport through a one dimensional time-dependent nanomechanical system. The microscopic model consists of coupled chains of atoms, considering local and non-local interactions between particles. We show that the system presents different stationary transport regimes depending on the driving frequency, temperature gradients and the degree of locality of the interactions. In one of these regimes, the system operates as a phonon refrigerator, and its cooling performance is analyzed. Based on a low frequency approach, we show that non-locality and its interplay with dissipation cause a decrease in cooling capacity. The results are obtained numerically by means of the Keldysh non-equilibrium Green's function formalism.
\end{abstract}


\end{frontmatter}

\section{Introduction}

Non-local interactions constitute the basis of a great variety of natural phenomena, at both macro and nanoscale levels, and heat transport is amongst them. Most of first principle studies on thermal transport in low dimensional systems, are based on classical and quantum models that consider local (first-neighbor) interactions \cite{nuestroPA,nuestroPA2, B. Ai,B. Hu, Freitas, G. Benenti, D. Segal,J. Behera}.

Regarding static systems, it was found that non-local/long-range (LR) interactions in a one-dimensional (1D) classical mass graded chain, connect distant particles with very different masses, enhancing system asymmetry and hence improving thermal rectification \cite {S.Chen}. It was also found in \cite {H.Zhou} that in 1D harmonic chains with different mass ordering, new conduction channels are opened forcing localized phonons to become delocalized, which results in an increase of thermal conductance.  On the other hand, in Ref.\cite{olivares} energy transport along a classical chain of unidirectionally aligned rotors was studied. The authors found that, while the system showed diffusive transport for short-range interactions, it behaved as a thermal insulator when the interaction range increased. The studies were also extended to anharmonic systems. In Ref.\cite {Bagchi,jwang} the effect of LR interactions in a Fermi-Pasta-Ulam (FPU) was analyzed, finding that the incorporation of non-local interactions increased thermal conductivity with respect to the local interaction case. Furthermore, it was found that thermal conductivity shows a strong dependence on interaction range.

The cited works highlight the role played by non-locality on heat transport properties. Interest in the role of non-local interactions in thermal transport through static systems is relatively recent, and its mechanism is not yet fully understood. The situation becomes less clear for time-dependent low dimensional systems, of which to our knowledge, there have not been any first principles studies devoted to explore non-local interactions and their effect on system´s thermal transport properties.

In our previous works \cite{nuestroPA,nuestroPA2} we treated the phononic heat transport through classical and quantum 1D chains with first-neighbor interactions between atoms, when subjected to thermo-mechanical time-dependent perturbations. We found that it is possible to {\it dynamically} tune  the presence of different transport regimes. In particular, we found that a system composed of two or three 1D quantum chains in contact through a time-dependent coupling, can act as a phononic refrigerator, pumping energy against a temperature gradient. The aim of the present work is to extend the study to the case of non-local interactions between particles of the same chain. We propose a one-dimensional quantum microscopic model subjected to mechanical time-dependent perturbations. The study is carried out numerically based on the Keldysh non-equilibrium Green function formalism \cite{M. Di Ventra}.

The article is organized as follows. Firstly, we present the model. Secondly, we obtain an expression for the stationary heat current. Thirdly, we discuss the transport regimes involved, comparing local and non-local cases, with a focus on phonon cooling and its performance. Finally, we present the main conclusions of this work.

\section{Model description}

\begin{figure}[ht]
\centering
\includegraphics[scale=0.089]{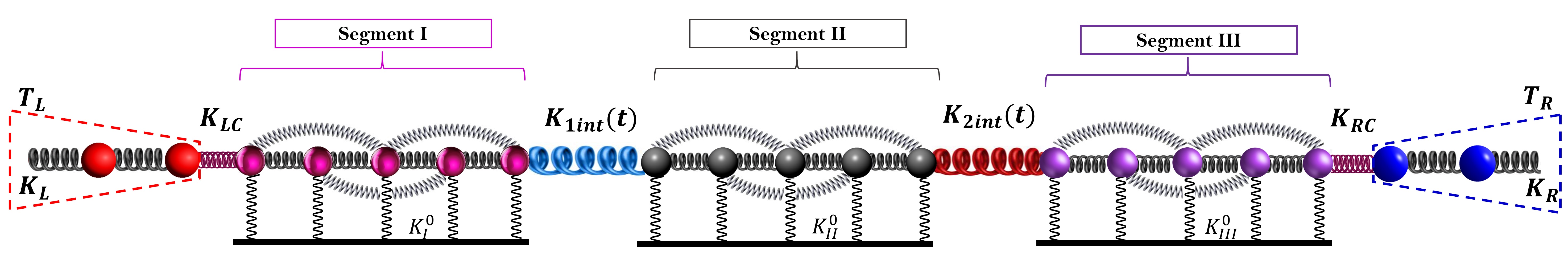} 
\caption{Sketch of the microscopic model.}
\label{fig:Figura1}
\end{figure}

We consider a one-dimensional chain of atoms, harmonically and bidirectionally coupled. This chain is made up of three segments (I, II and III). The central chain (II) is coupled to the other two segments (I and III) through a time-dependent mechanical interaction (see sketch in Fig.\ref{fig:Figura1}).

We assume that the particles can interact locally (first-neighbor) and non-locally (second-neighbor) and can vibrate only longitudinally. The atoms are also subject to a harmonic pinning potential that models the interaction of the system with a substrate. In turn, its left (L) and right (R) ends are connected to two macroscopic systems formed by semi-infinite chains of atoms with masses $m_L$ and $m_R$ and harmonically coupled with elastic constants $K_L$ and $K_R$. This macro-systems are in thermal equilibrium at temperatures $T_L$ and $T_R$ respectively and play the role of thermal reservoirs  (Rubin model \cite{rubin}). Atoms within each of the three segments interacts to first-neighbor (local) o second-neighbor (non-local).
The Hamiltonian of the system is:
\begin{equation}
H(t)=H_{central}+H_{int}(t)+H_{contact}+H_{\beta},
\end{equation}
where $H_{central}$ describe the contribution of segments $(\alpha= I, II , III)$, $H_{int}$ the interaction between segments $I-II$ and $II-III$,  $H_{contact}$ the contact between the central chain and the reservoirs $L,R$ and $H_{\beta}$ the contribution of the reservoirs $\beta= L, R$.
\begin{equation}
H_{central}=\sum_{\alpha} \sum_{i=1}^{N_{\alpha}} \frac{p_{i,\alpha}^2}{2m_{i,\alpha}}+\\
\sum_{j=1,j\neq i}^{N_{\alpha}} \sum_{i=1}^{N_{\alpha}}\frac{1}{2} K_{i,j}^{\alpha,\nu}(x_{j,\alpha}-x_{i,\alpha})^2+
\sum_{i=1}^{N_{\alpha}-1} \frac{1}{2} K_{\alpha}^0 x_{i,\alpha}^2,
\end{equation}
with $N_{\alpha}=\frac{N}{4},\frac{N}{2},\frac{N}{4}$ is the length of the $I,II,III$ segment, respectively. $m_{i,\alpha}$ is the mass $i-th$ of the atom, $x_{i,\alpha}$ and $p_{i,\alpha}$ represent the displacement from its equilibrium position and momentum respectively of the i$-th$ atom of the $\alpha$ segment.
$K_{i,j}^{\alpha,\nu}=\frac{K_{\alpha}}{|i-j|^\nu}$ are the elastic constants between the i$-th$ and j$-th$ atoms of the $\alpha$ segment, with $K_{\alpha}$ a reference value. $K_{\alpha}^0$ is the strength of the pinning potential, $N_{\alpha}$ is the number of particles in the segment $\alpha$ and $\nu$ the range of the interaction. $\nu=0$ corresponds to mean field limit and $\nu \rightarrow \infty$ corresponds to first-neighbor (local) interaction.  In this work we consider $\nu=2$, so the strength constant for second neighbors interactions $(k-k'=2)$ is $25\%$ of the local case. As we are interested in the heat transport through small size systems, this assumption is a quite good approximation to study non-local interactions.

On the other hand, the interactions between structures and with the reservoirs are local (first neighbors)(see Fig.\Ref{fig:Figura1}). The Hamiltonian of the interaction can be written as: 
\begin{equation}
H_{int}= \frac{1}{2}K_{1,int}(t)(x_{N_I-1,I}-x_{1,II})^2+ \frac{1}{2}K_{2,int}(t)(x_{N_{II}-1,II}-x_{1,III})^2,
\end{equation}
with 
\begin{eqnarray}
\label{ec.acoplamiento1}
K_{1,int}(t)=K^0_{1,int}(1+\epsilon \cos(\omega_0 t)),\\
K_{2,int}(t)=K^0_{2,int}(1+\epsilon \cos(\omega_0 t+\phi)),
\label{ec.acoplamiento2}
\end{eqnarray}
where $K_{1,int}(t)$ and $K_{2,int}(t)$ are the time-modulated coupling constants between the segments $I$-$II$ and $II$-$III$ respectively, and oscillate out of phase, with a phase difference $\phi$. The time-dependent perturbations produce a temporal symmetry breaking.
The contact with the thermal reservoirs is described by the Hamiltonian,
\begin{equation}
H_{contact}=\frac{1}{2}K_{L_c}(x_{1,L}-x_{1,I})^2+\frac{1}{2}K_{R_c}(x_{1,R}-x_{N_{III}-1,III})^2,
\end{equation}
where elements $(1,I)$ and $(N_{III}-1,III)$ are the atoms of the system in contact with the first atoms (``1'') of the reservoirs.
Finally, the Hamiltonian of the thermal reservoirs $\beta$ can be written in therms of their normal modes (see Appendix I)

\begin{equation}
\label{res_normal}
H_{\beta}= \sum_{k_{\beta}=0}^{N_\beta} \frac{p_{k\beta}^2}{2 m_{k\beta}}+\frac{1}{2} K_{\beta}[1- \cos(u_{k_{\beta}}) ] x_{k_{\beta}}^2, \end{equation}
where $k_{\beta}$ is the $k$ mode of the $\beta$ reservoir.
For the numerical simulations we use dimensionless parameters: spring constants $K_i$ in units of $K_R$, positions in units $[a]$, moments in units $[a(m K_R)^{1/2}]$, frequencies in units $[(K_R/m)^{1/2}]$ and temperatures in $[a^2K_R/k_B]$. For a typical atom, $a \sim 0.1$nm and mass $m\sim 10^{-26}$kg. On the other hand, we consider typical frequencies $\omega \sim 10-100$GHz. that are smaller than the inverse of typical electron–phonon relaxation times $\thicksim 0.1$ps (e.g. Si), in order to consider only the relevant time scales of the phonon scattering processes. The reference temperature is $T_0 \sim 50$K  that is smaller than typical Debye temperatures. 

We define the force matrix $F(t)$ of the system as, $F(t) =F^{0}(t)+F^{0(NL)}(t)+Fo^{0}(t)+F^{1}(t)$ where $ F^{0}(t)$ includes the local inter particle forces and the system-reservoir interaction, $F^{0(NL)}(t)$ represent the non-local inter particle forces within a segment.  $Fo^{0}(t)$ refers to the on-site interaction and $F^{1}(t)$ accounts for the time-dependent contribution to the force. The complete expression of the force matrix can be found in the Appendix II.

From the continuity equation and energy conservation, the local time-dependent heat current from the site $l$ of the central chain $\alpha$ to each reservoir $\beta$ can be expressed as:
\begin{equation}
\label{curR}
J_{\beta}(t)= \sum_{l_{\alpha}} \frac{K_{\beta C}}{m_\alpha} \langle  x_{{\beta}} p_{l_{\alpha}}\rangle.
\end{equation}

The heat flux is defined positive when enters the reservoir. The position of each atom of the semi-infinite chains (reservoir) can be expressed in terms of the normal modes. Therefore, the heat current given in Eq. \ref{curR} can be written as:
\begin{equation}
\label{curRR}
J_{\beta}(t)= \sum_{k_\beta, l_{\alpha}} \gamma_{k_{\beta},l_{\alpha}} \langle  x_{k_{\beta}} p_{l_{\alpha}}\rangle, \end{equation}
where $l_\alpha$ refers to the site of the system connected to the $\beta$ reservoir and $k_{\beta}$ is the normal mode $k$ of reservoir $\beta$. $\gamma_{k_\beta,l_{\alpha}}$ are the coupling parameters, that can be written in terms of the coupling constant and the amplitudes of the normal modes (see Appendix I, Eq. \ref{mnR}-\ref{eqnacoplam}).

To evaluate the $DC$ heat current we use the Keldysh non equilibrium Green formalism. We define the retarded ($^R$), lesser ($^>$) and greater ($^<$) Green functions, that can be expressed as
\begin{equation}
G^{<}_{l_{\alpha}, k_{\beta}} (t,t')= \textit{i}  \left\langle x_{k_{\beta}}(t') x_{l_{\alpha}}(t)\right\rangle,
\end{equation}
\begin{equation}
G^{<}_{k_{\beta}, l_{\alpha}}(t,t') = \textit{i}  \left\langle x_{l_{\alpha}}(t') x_{k_{\beta}}(t)\right\rangle,
\end{equation}
\begin{align*}
      G^{R}_{k_\alpha,k'_\alpha}(t,t')&= -\textit{i} \Theta(t-t') \langle [x^\dag_{k_\alpha}(t'),x_{k_\alpha}(t') ] \rangle,  \\
     & = \textit{i}\Theta(t-t') [G^{<}_{k_\alpha,k'_\alpha}(t,t')-
G^{>}_{k_\alpha,k'_\alpha}(t,t')],
\end{align*}
with $x_{k}(t)$ a phononic operator \cite{AMCC}.
To calculate the equations of motion of $G^{R}_{k_\alpha, k'_\alpha}(t,t')$ we take the second derivative with respect to $t'$ and use the Ehrenfest's theorem, following the strategy described in Ref.\cite{AR} obtaining:
\begin{multline}
\label{eq:eheren}
-\left[ \partial^{2}_{t'}+F_{k_\alpha,k'_\alpha} \right] G^{R}_{k_\alpha, k'_\alpha} (t,t')+F_{k_\alpha,k'_\alpha+1} G^{R}_{k_\alpha, k'_{\alpha}+1} (t,t') + F_{k_\alpha,k'_\alpha-1} G^{R}_{k_\alpha, k'_\alpha-1} (t,t')=\\
=\frac{1}{m_\alpha}\delta_{k_\alpha,k'_\alpha} \delta (t-t')+
F_{k_\alpha,l_{\alpha}} G^{R}_{k_\alpha, l_{\alpha}} (t,t')\delta_{k'_\alpha,l_{\alpha}}-
F_{k_\alpha,l_{\beta}} G^{R}_{k_\alpha, l_{\alpha}} (t,t')\delta_{k'_\alpha,l_{\alpha}},
\end{multline}

Integrating the degrees of freedom of the reservoirs in the Dyson equation, Ref. \cite{J.-S. Wang,jauho} for the retarded Green's function along the contacts we obtain
\begin{equation}
\label{DysonEq}
-\partial^2_t \hat{G}^R(t,t') + \hat{G}^R(t,t')\hat{F}(t')-\int dt_1 \hat{G}^R(t,t_1)
\hat{\Sigma}^R(t_1,t')=\delta(t,t') \mathcal{I}.
\end{equation}
Representing the time-dependent perturbation of the Hamiltonian in terms of its Fourier expansion $ F^{(1)}(t)=\sum_{k=-1, \neq 0}^{k=1} \hat{F}_k^{(1)} e^{i k \omega_0t}$, substituting in the Dyson's equation (Eq. \ref{DysonEq}) and performing the Fourier transform with respect to the ``delayed time'' $t'$ in $G^R(t,t')$, results

\begin{equation}
\label{GreenretardT}
\hat{G}^R(t,\omega)=\hat{G}^{\left(0\right)}(\omega)+\sum^K_{\substack{k\neq 0}} e^{-ik \omega_{0} t} \hat{G}^R \left(t,\omega+k\omega_{0} \right)
\mathcal{\hat F}^{(1)}_{k} \hat{G}^{(0)}(\omega).
\end{equation}
Generally this system of equations (Eq. \ref{GreenretardT})  can be solved numerically.  However, it may be convenient to carry out systematic expansions of $H_ {int}$ in powers of $ \hat {F} ^ {(1)} $ in order to obtain analytical expressions. Considering a perturbative approach, that is, if $H_{int}(t)$ small compared to the time-independent Hamiltonian, a low-order expansion can be a good approximation. As $G^R(t,\omega)$ has the same temporal periodicity of $H(t)$, it is possible to expand it in terms of the Floquet components  $\hat{G}(k,\omega)$
\begin{equation}
\label{Greenretardfloquet}
\hat G^{R}(t,\omega)=\sum_{k=-\infty}^{\infty} e^{-i k \omega_{0} t} \hat{\mathcal{G}}(k,\omega).
\end{equation}

\begin{equation}
\label{Greenretardfloquet2}
\hat{G}^{R}(t,\omega)\approx \sum_{k'=-K}^{K} e^{-i k' \omega_{o} t} \hat{\mathcal{G}}(k',\omega),
\end{equation}
with
\begin{align}
\label{gocompl}
\hat{\mathcal{G}}(0,\omega)&=\hat{G}^0(\omega)+\sum_{k\neq 0} \hat{G}^0(\omega) \hat{\mathcal{F}}_{-k}^{(1)} \; \hat{\mathbb{G}}(k,\omega),\\
\label{g1compl}
\hat{\mathcal{G}}(\pm k,\omega)&=\hat{G}^0(\omega\pm k \omega_{0})[\hat{\mathcal{F}}^{(1)}_{\pm k}\;  \hat{G}^{0}(\omega)+\sum_{k'\neq 0} \hat{\mathcal{F}}_{\pm k-k'}^{(1)} \hat{\mathbb{ G}}(k',\omega)],
\end{align}

where it has been defined  $\hat{\mathbb{ G}}(k,\omega)=\hat{G}^0(\omega+k \omega_0)\hat{\mathcal{F}}_k^{(1)}\hat{G}^0(k,\omega)$. 

Since in our study the time-dependent perturbation contains only one harmonic, the summations will contain only the terms $ k = \pm 1 $. Moreover, we only consider processes that involve the exchange of a single phonon $ \omega_0 $ and very weak disturbances, therefore expressions in Eqs. \ref{gocompl} and \ref{g1compl} can be written as:
\begin{align}
\label{ec.intercambio}
\hat{\mathcal{G}}(0,\omega)&=\hat{G}^0(\omega),\\
\label{ec.intercambio1}
\hat{\mathcal{G}}(\pm 1,\omega)&=\hat{G}^0(\omega\pm \omega_{0})\hat{F}^{1}_{\pm 1} \hat{G}^{0}(\omega),
\end{align}

with
\begin{equation}
    \label{Go}
    \hat{G}^0(\omega)[\omega ^2 \mathcal{I}-{\mathcal{\hat F}}^0-\hat \Sigma^R(\omega)]=\mathcal{I},
\end{equation}
the stationary component of the retarded Green's function of the central chain connected to reservoirs with no time-dependent perturbation.

$\Sigma^{R}(\omega)$ is the self energy defined in Appendix III.

In Fig.\ref{fig:intercambiofonones} we depict different processes involving only an exchange of energy with the reservoirs corresponding to absorption/release of one $\hbar \omega_0$ phonon.
\begin{figure}[H]
    \centering
    \includegraphics[scale=0.25]{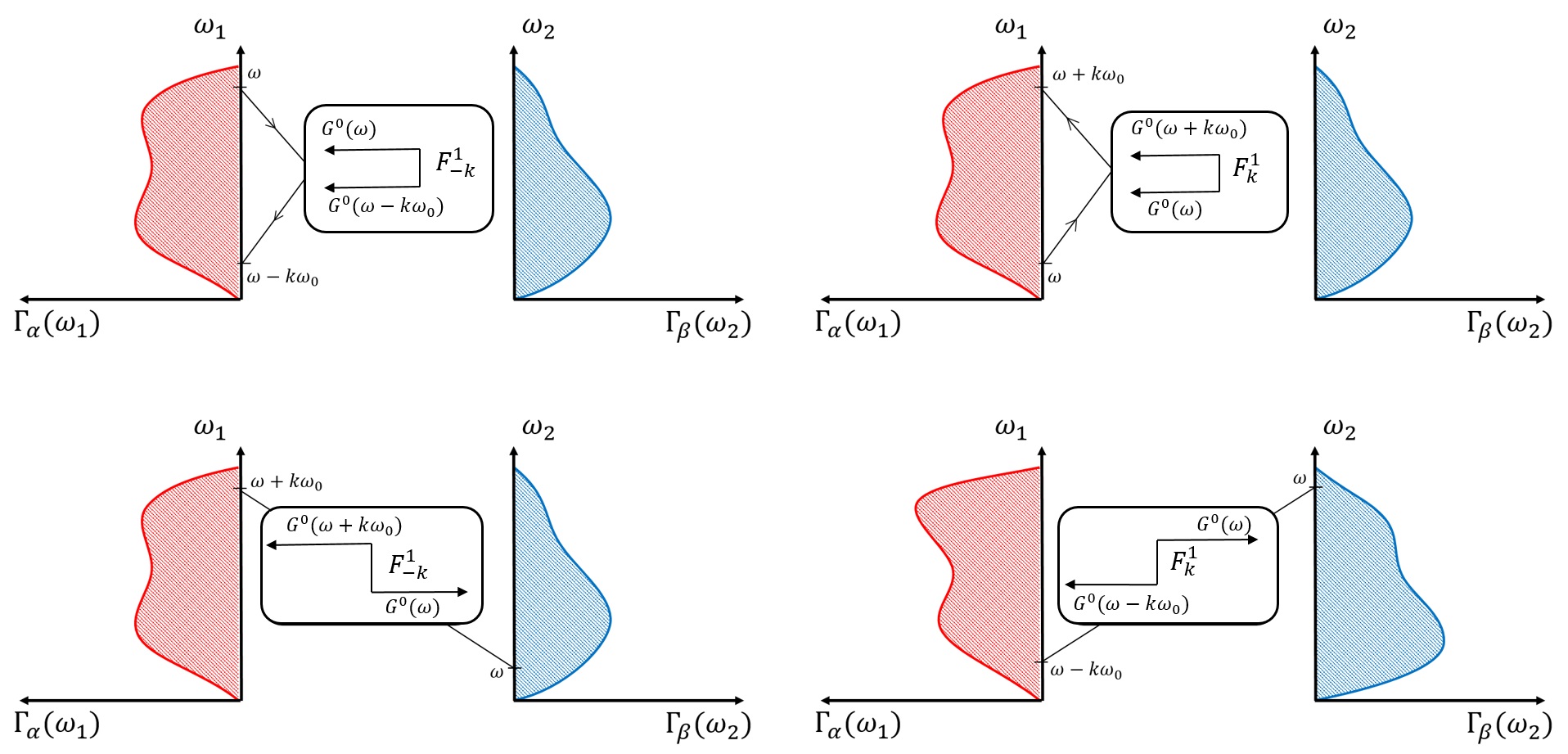}
    \caption{Processes of exchanges of energy between reservoirs $\alpha$ and $\beta$. }
    \label{fig:intercambiofonones}
\end{figure}

Conservation of energy implies that the total average power $\bar{P}$ released by the external drivings during a time period equals the energy dissipated into the reservoirs at a rate
$\sum_{i=L,R} \bar{J}_i=\sum_{i=1}^2\bar{P}_i$, with $\bar{J}_{\beta}$ the stationary heat current flowing in/out of reservoir $\beta$.

From the definition of the heat current in Eq. \Ref{curRR} we can rewrite $J_{\beta}(t)$ as
\begin{equation}
 J_{\beta}(t)=-\sum_{k_{\beta},l_{\alpha}}\frac{K_{\beta c}}{m_{\alpha}}\sqrt{ \frac{2}{(N_{\beta} +1)}}\sin(u_{k_\beta})\lim_{t\rightarrow t'}
\Re\left[i\frac{\partial}{\partial t'} \left\langle x_{k_{\beta}}(t) x_{l_{\alpha}}(t')\right\rangle\right],
\label{currentgreen} 
\end{equation}
that can be expressed in terms of the lesser Green function as
\begin{equation}
 J_{\beta}(t)=-\sum_{k_{\beta},l_{\alpha}}\frac{K_{\beta c}}{m_{\alpha}}\sqrt{ \frac{2}{(N_{\beta} +1)}}\sin(u_{k_\beta})\lim_{t\rightarrow t'}
\Re\left[i\frac{\partial}{\partial t'}G^{<}_{l_{\alpha}, k_{\beta}}(t,t')\right ].
\label{currentgreen2} 
\end{equation}

Following a similar procedure described in \cite{nuestroPA,AMCC}, the $DC$ heat current flowing in/out of the reservoir $\beta$ can be written as:
\begin{multline}
   \label{fullJ}
\bar {J}_{\beta}=\sum_{{\beta}'=L,R}\sum_{\substack{k=-1}}^{1}\int_{-\infty}^{\infty}\frac{d\omega}{2\pi} (\omega+k\omega_{o})\left[\eta_{\beta'}(\omega)-\eta_{\beta}(\omega+k\omega_{o})\right]\times\\
\times\Gamma_{\beta}(\omega+k\omega_{o}) \Gamma_{\beta'}(\omega)\left|{\mathcal{G}}_{l_{\alpha}, l_{\alpha'}}(k,\omega)\right|^2,
\end{multline}
where $\Gamma_\beta(\omega)$ and $\eta_\beta$ are the spectral density and the Bose-Einstein distribution of the reservoir $\beta$ respectively and $\eta_\beta(\omega)$ is defined as $\eta_\beta(\omega)=1/({e^{\frac{\hbar \omega}{k_B T_\beta}}-1})$, $k_B$ is the Boltzmann constant.

In the steady state, the value of the current in each segment is independent of the site $i$, therefore $\bar J_{i,I} = \bar J_L$ and $\bar J_{i,III} = \bar J_R$.

\section{Results}

\begin{figure}[ht]
\centering
\includegraphics[scale=0.55]{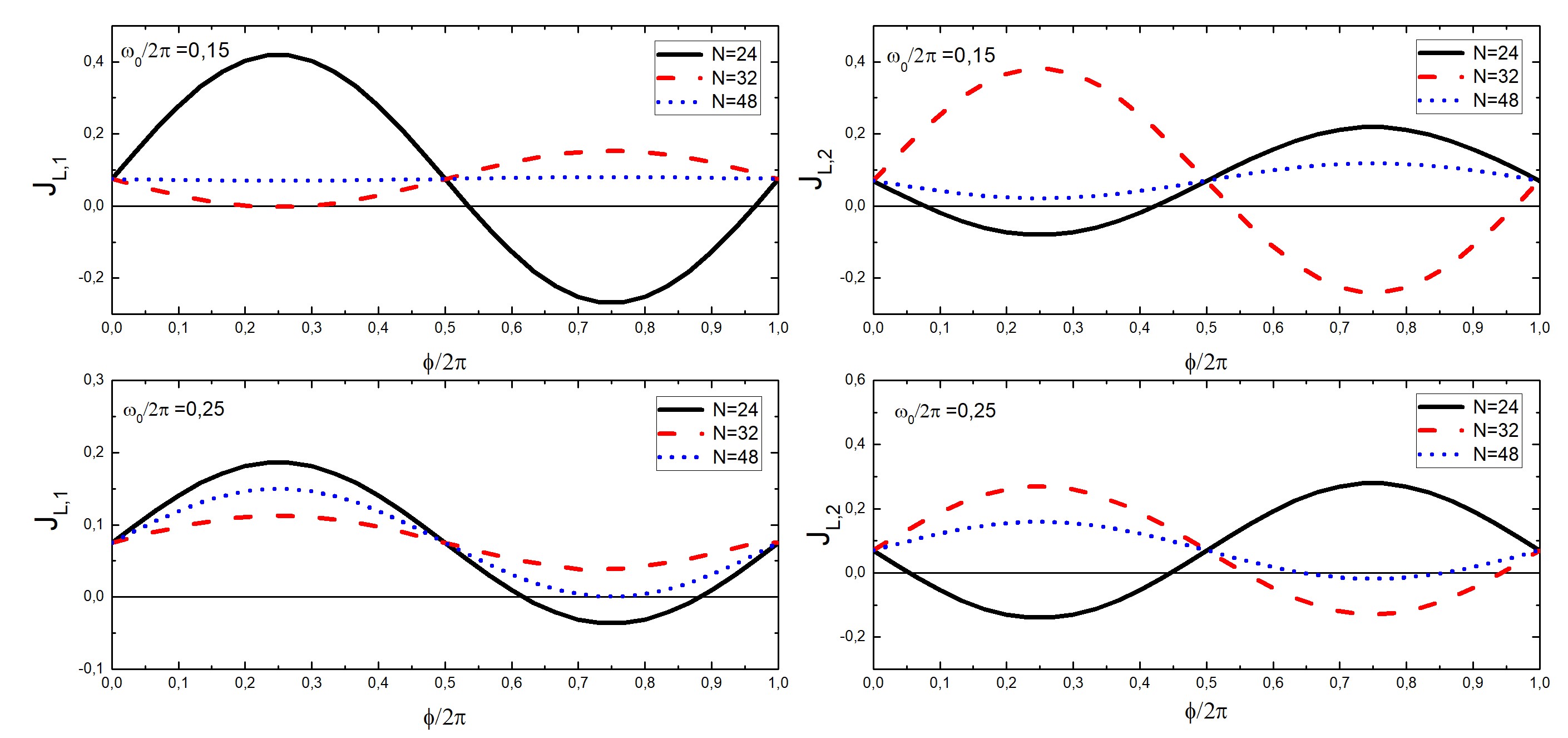}
\caption{$J_{L,1}$ and $J_{L,2}$ vs $\phi /2\pi$, where the subscript 1, 2 represent the local and non-local interaction respectively. $\omega_0/2\pi=0.15 ; 0.25$ $T_L=1.0$ y $T_R=1.1.$ 
The line patterns, black solid line, red dashed line, blue dotted line, represent N=24, 32, 48 respectively. }
\label{fig:JLvsfi}
\end{figure}

In order to analyze the existence of different transport regimes, we calculate the $DC$ current $J_{\beta}$ given in Eq. \ref{fullJ} as a function of the dynamical parameters that characterized the temporal modulation. In Fig. \Ref{fig:JLvsfi} we plot $J_L$ versus $\phi$ for different values of $\omega_0$ and for different system sizes.

The left and right columns of Fig. \Ref{fig:JLvsfi} correspond to local and non-local ($1st$ and $2nd$ neighbors) interactions, respectively. It can be observed that in both cases, $J$ depends non monotonically on the phase shift $\phi$, and that the current profiles oscillate around a mean value of $J_L$ corresponding to the current for the static case at a given temperature gradient.
However, different behaviors can be found depending on the value of $\omega_0/2\pi$ and $N$.
For example in the case of $N=32$ (red dashed curve) and local interactions, $J_L$ is zero only for a single value of frequency: $\omega_0/2\pi=0.15$. However, when the interaction is non-local, the heat flow towards the reservoir $L$ is zero for different values of $\omega_0$ and $\phi$, so the system works as a local insulator. The case $N=24$ (black solid curve) presents another interesting behavior. Local insulation regimen is observed again for both local and non-local interaction, but in this specific case the non-locality can produce an inversion in current direction. For case $N=48$ (blue dotted curve) local interaction produces a current as in the static case for every value of $\phi$, that's mean the system behaves as time independent, i.e. the net power released into the system is zero. 
We present arbitrarily results for $J_L$ to discuss, but we remark that $J_R$ presents qualitative similar regimes (not shown here).

Temperature gradient and its interplay with the time-dependent modulation results as other source of non-equilibrium that affects the thermal transport. The expression of the $DC$ current given in Eq. \ref{fullJ} points out to the interdependence between frequency, phase shift, temperatures and the degree of locality encoded in the Green's functions.

In a previous work \cite{nuestroPA} we showed that with first neighbor interactions and according to the temperature differences $\Delta T = T_R-T_L$ different transport regimes regarding the operational mode of the system can be found: thermal rectifier, heat engine, thermal sink and refrigerator (pumping energy against a temperature gradient).

\begin{figure}[ht]
\centering
\includegraphics[scale=0.75]{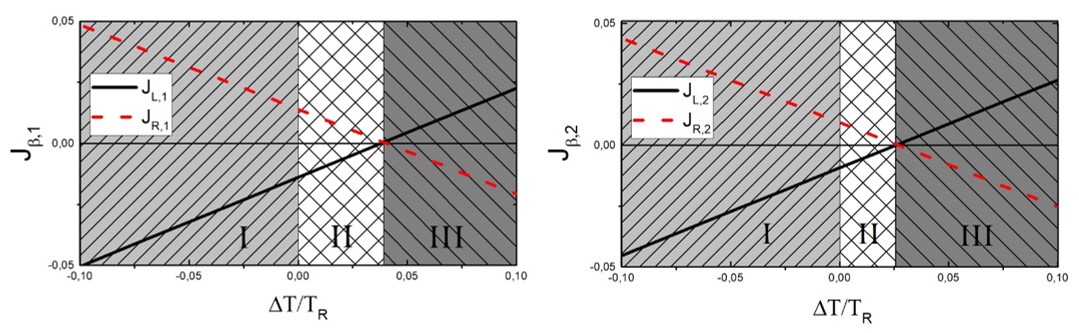}
\caption{ $J$ vs. $\Delta T/T_R$ with $\Delta T=(T_R-T_L)$ and $T_R=0.1$; for $N=24$ and $\frac{\omega_{0}}{2 \pi}=0.15$ and $\frac{\phi}{2 \pi}= 0.8$. Left/right panels correspond to local/ non-local interactions, respectively. The curves refer to $J_L$(black solid line)$/J_R$(red dashed line). The constants are: $K_{I}^0=45.0; K_{II}^0=40.0; K_{III}^0=50.0; K_{1int}^0=1.05; K_{2int}^0=1.05; K_{LC} = 20.1; K_{RC} = 1; K_L = 40.5; K_R = 5.5$.}
\label{fig:J_vs_deltaT_N24}
\end{figure}

In this work, we search to these regimes and how are affected by the non-local interactions. We compare the local and non-local cases (up to second neighbors), considering short chains to enhance the effect of the non locality. In Fig. \ref{fig:J_vs_deltaT_N24} we plot $J_\beta$ as a function of $\Delta T$ for $N=24$, where $T_R$ is set as the reference temperature. The black solid/red dashed curves represent $J_L$ and $J_R$, respectively.

For both local and non-local interactions, the current displays a linear dependence with $\Delta T$. It is interesting to note that,  unlike what happens in a static system, if $\Delta T = 0$ the heat current is non-zero in both segments. Therefore, there is a net heat current assisted by the phonon pumping. 
Moreover, the system rectifies the heat current, that is,  $|J_L| \neq |J_R|$ when the sign of the temperature gradient is inverse. This rectification phenomena are of a {\it dynamical} nature.

\begin{figure}[ht]
\centering
\includegraphics[scale=0.3]{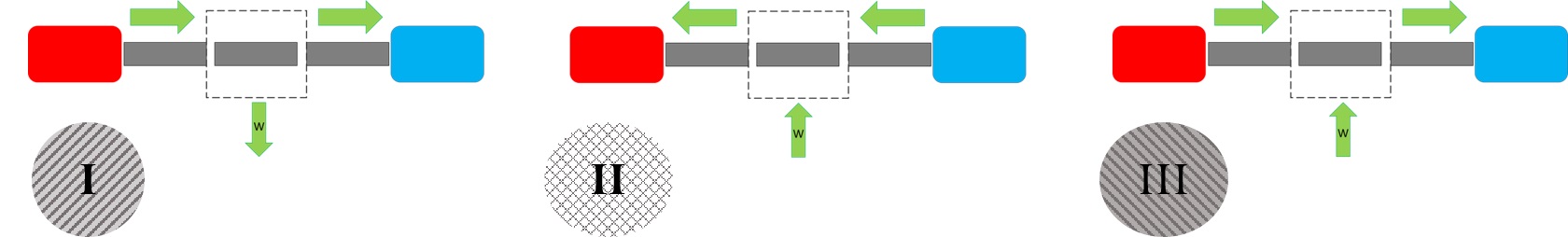}
\caption{Diagrams of the different heat-work flow circuits corresponding to the transport regimes indicated in Fig.\ref{fig:J_vs_deltaT_N24}. }
\label{fig:flujos}
\end{figure}

In Fig. \ref{fig:J_vs_deltaT_N24} shaded regions indicate different regimes $(I,II,III)$. Regions $I$ and $III$ correspond to values of $\Delta T$ in which the heat flows from high to low temperatures, represented by the schemes $I$ and $III$ in Fig. \ref{fig:flujos}. The only difference between these two cases is that in scheme $I$/$III$ the power is injected to/extracted from the system respectively.

On the other hand, the central region $II$ indicates a regime in which both $J_L$ and $J_R$ have a direction against the temperature gradient. This situation corresponds to the cooling regime sketched as $II$ in Fig. \ref{fig:flujos}. However, the range of $\Delta T$ for which the system acts as a refrigerator (the white band - region $II$), depends on the locality, becoming wider for the non-local case. 

It is important to note that regardless the type of interaction there are values of $\Delta T$ for which both currents are zero, thereby the system behaves as a thermal insulator and this behavior is found for chains with different lengths.

\begin{figure}[ht]
 \centering
\includegraphics[scale= 0.5]{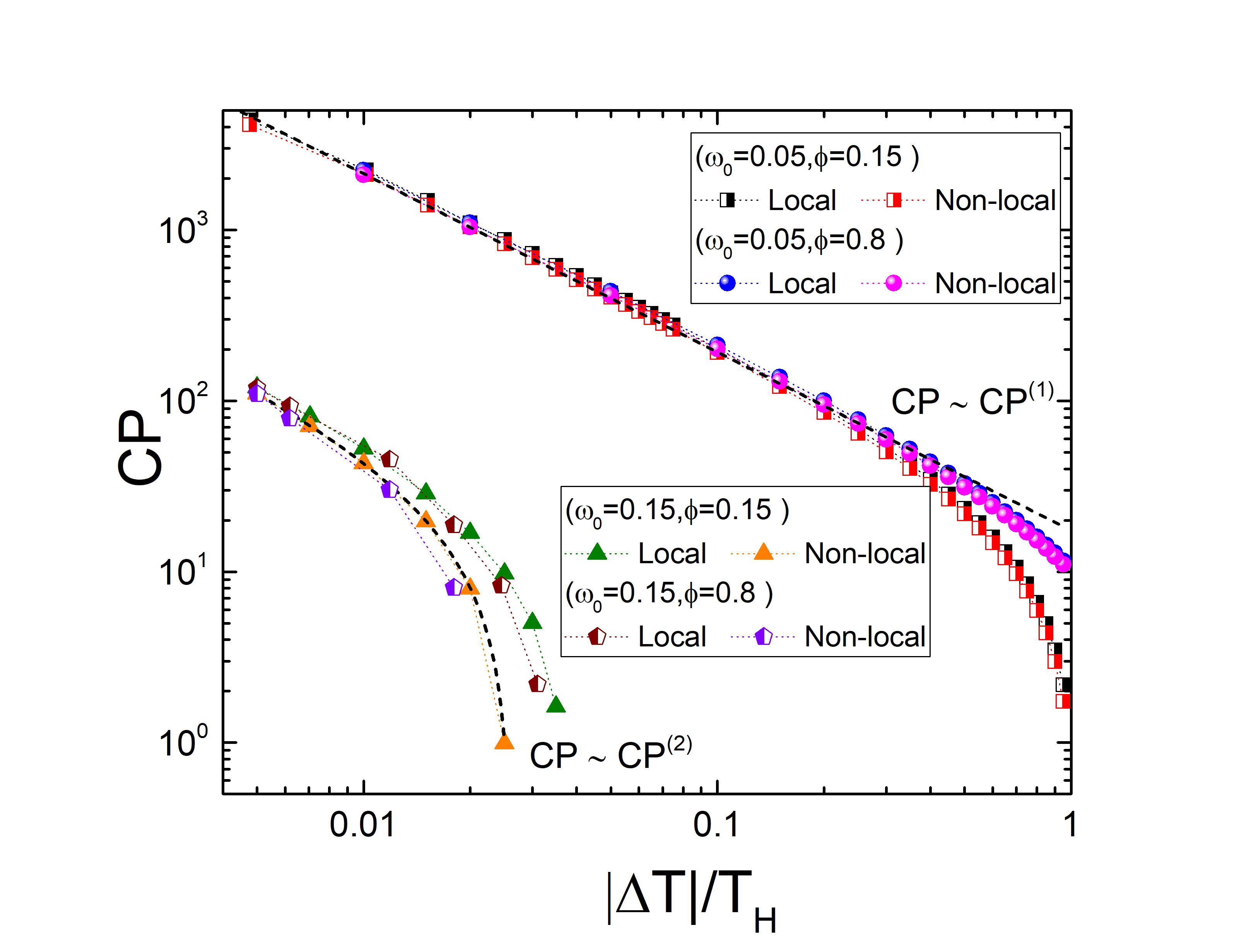}
   \caption{Cooling performance coefficient $CP$ vs $\Delta T/T_H$ with $\Delta T=|T-T_H|$ and $T_H=0.1$, for the case of local ($j=1$) and non-local ($j=2$) interactions, and $N=24$. a), b), c) and d) correspond respectively to the following set of parameters ($\omega_{0},\phi)={(0.05;0.15),(0.05;0.8),(0.15;0.15),(0.15;0.8)}$. $CP^{(1)}(\Delta T)$ and $CP^{(2)}(\Delta T)$ are plotted with dashed lines for $(\omega_{0},\phi)=(0.05;0.15)$ and $(0.15;0.15)$ respectively. }
      \label{fig:CPvsDT}
\end{figure}

To characterize the cooling regime, we define the cooling performance coefficient $CP$ as
\begin{equation}
\label{CP}
CP = \frac{|\dot Q_{C}|}{|\dot Q_{H}+\dot Q_{C}|}
\end{equation}
where $\dot Q_{C}$ and $\dot Q_H$  are the heat currents flowing out/into the cold/hot reservoirs respectively, and $\dot Q_{H}+\dot Q_{C}$ the power developed by the external forces.
In Fig. \ref{fig:CPvsDT} $CP$ is plotted as a function of the $\Delta T$ normalized to the temperature of the hot reservoir ($T_H$) and  considering local and non-local interactions. 

We find that $CP$ decreases with $\Delta T/T_H$ for both local and non-local cases and for all parameter sets considered. When $\Delta T/T_H$ increases, $CP$ decreases up to a minimum value that corresponds to the maximum $\Delta T/T_H$ for which the system acts as a refrigerator.
Besides, we find that $CP$ for non-local interactions takes equal or smaller values than local case, sweeping in both cases several orders of magnitude, as can be seen in Fig. \ref{fig:CPvsDT}. 
On the other hand, we observe curves with similar locality and frequency, but different phase shift separate each other, an effect that is enhanced as long as $\Delta T$ increases.
We can be shown that for small $\Delta T/T_H$, $CP$ depends on $\Delta T$ according to a power law: $CP \propto  {\Delta T}^{-\alpha}$, with $\alpha$ a characteristic exponent that depends on the frequency. In the next section, we present an adiabatic analysis of the $DC$ heat current to investigate the origin of this scaling law.

\subsubsection*{Low frequencies heat current expansion $\omega_0$}

In order to analyze the mechanism underlying the scaling law of $CP$ with $\Delta T$, we expand the $DC$ heat current $J_\beta$, defined in Eq. \ref{fullJ}, up to second-order in $\omega_0$. In the low-frequency limit, the spectral density $\Gamma(\omega)$ and the Bose-Einstein distribution $n_\beta(\omega)$ can be approximated as
\begin{eqnarray}
\Gamma_L(\omega+ k\omega_0)\approxeq&\Gamma_L(\omega)+k \omega_0 \frac{d\Gamma_L(\omega)}{d\omega},\\
\eta_L(\omega+k\omega_0)\approxeq& \eta_L(\omega)+\frac{\partial \eta_L(\omega)}{\partial \omega} k \omega_0.
\end{eqnarray}
The difference between the Bose-Einstein distributions of each reservoir can be expressed as 
\begin{equation}
\label{difbosse}
    \eta_R(\omega)- \eta_L(\omega+k\omega_0)\approxeq\frac{\partial n_L}{\partial T}\Delta T- \frac{\partial n_L}{\partial \omega} k \omega_0.
\end{equation}
Given that $\partial_{T} \eta_L(\omega)=\left(-\frac{\omega}{T}\right) \partial_\omega n_L(\omega)$, the  Eq. \ref{difbosse} can be rewritten as
\begin{equation}
    \eta_R(\omega)-\eta_L(\omega+k\omega_0)\thickapprox\frac{\partial \eta_L(\omega)}{\partial\omega}\left[-\frac{\omega}{T_L}\Delta T+k \omega_0\right].
\end{equation}
replacing these approximations in Eq. \ref{fullJ}, $J_\beta$ in the adiabatic limit can be expressed as
\begin{eqnarray}
\label{ec:JL}
    J_\beta&\approx&\sum_{\beta'=R,L} \sum_k\int\frac{d\omega}{2 \pi}(\omega+k\omega_0)    \left[\frac{\partial \eta_\beta}{\partial \omega} \left( -\frac{\omega}{T_\beta}\Delta T-k\omega_0\right)\right]\Gamma_{\beta'}(\omega)\times \nonumber \\ 
    &&\times \left[ \Gamma_\beta(\omega)+k \omega_0 \frac{d\Gamma_\beta(\omega)}{d\omega}\right] {\arrowvert \mathcal{G}_{f_{l \alpha,l \alpha'}} (k,\omega)\arrowvert}^2,
\end{eqnarray}
where $\mathcal{G}_{f} (k,\omega)$ is the Floquet component of the  {\it frozen} Green function $\hat{G}_{f} (t,\omega)$ defined as
\begin{equation}
\label{dftw}
    \hat{G}_f(t,\omega)=\left[ \hat{G}^{(0)}(\omega)^{-1} - \hat F^{(1)}(t)\right]^{-1},
\end{equation}
with ${F}^{(1)}(t)$ being the component of the force matrix related to the time-dependent perturbation. This {\it frozen} Green function accounts for a regime in which the dynamical matrix adjusts instantaneously to the AC mechanical perturbation \cite{N. Bode} and can be obtained from the Dyson equation (Eq. \ref{DysonEq}) in the low frequency limit. An exact solution up to order $\omega_0$ can be obtained by expanding Eq. \ref{GreenretardT} as: 
\begin{equation}
\hat{G}(t,\omega)\sim {\hat G}^{(0)}(\omega)+{\hat G}(t,\omega) \hat F^{(1)}(t) {\hat G}^{(0)}(\omega)+i \partial_{\omega} {\hat G}(t,\omega) \frac{d\hat F(t)}{dt} {\hat G}^{(0)}(\omega).
\end{equation}
Considering the definition given in Eq. \ref{dftw}, the exact solution of the Dyson equation at order $\omega_0$ reads
\begin{equation}
\hat G^{(1)}(t,\omega)= {\hat G}_{f}(t,\omega)+i \partial_{\omega}{\hat G}_{f}(t,\omega) \frac{d\hat F^{(1)}(t)}{dt}{\hat G}^{(0)}(\omega).
\end{equation}
Using all these approximations and expanding Eq. \Ref{ec:JL} in terms of $\omega_0^0$, $\omega_0^1$ and $\omega_0^2$ and neglecting higher order terms $O(\omega_0^3)$, the current $J_\beta$ becomes
\begin{equation}
\label{Ec.jalphaaprox}
    J_\beta\approx J_\beta^{(0)}+ J_\beta^{(1)}+J_\beta^{(2)},
\end{equation}
$J_\beta^0$ is the zero-order component, and corresponds to the purely thermal contribution due to the temperature gradient.
\begin{equation}
    \label{ec:JLordcero}
    J_\beta^{(0)}\approx\sum_{\beta'} \int\frac{d\omega}{2 \pi}\left(-\frac{d n_\beta}{d\omega}\right) \frac{\omega^2 \Delta T}{T_\beta}\Gamma_{\beta'}(\omega) \Gamma_\beta(\omega) |\mathcal{G}_{f_{l\alpha,l \alpha'}}{(0,\omega)}|^2.
\end{equation}
The second term in expansion Eq. \ref{Ec.jalphaaprox} corresponds to the contribution to first order in $\omega_0 $ and can be written as
\begin{equation}
 \begin{aligned}
    \label{ec:JLorduno}
       J_\beta^{(1)}\approx  &\sum_{\beta'} \sum_k\int\frac{d\omega}{2 \pi}
        \left(-\frac{d \eta_\beta}{d\omega}\right) \frac{\Delta T }{T_\beta}\omega k\omega_0 \Gamma_{\beta'}(\omega) \Gamma_\beta(\omega) |\mathcal{G}_{f_{l\alpha,l \alpha'}}{(k,\omega)}|^2+\\
        &+\sum_{\beta'} \sum_k\int\frac{d\omega}{2 \pi}
        \left(+\frac{d \eta_\beta}{d\omega}\right) k\omega_0 \omega \Gamma_{\beta'}(\omega) \Gamma_\beta(\omega) |\mathcal{G}_{f_{l\alpha,l \alpha'}}{(k,\omega)}|^2+\\
        &+\sum_{\beta'} \sum_k\int\frac{d\omega}{2 \pi}
        \left(-\frac{d \eta_\beta}{d\omega}\right) \omega^2\frac{\Delta T}{T_\beta} \Gamma_{\beta'}(\omega) k \omega_0 \frac{d\Gamma_\beta }{d\omega} |\mathcal{G}_{f_{l\alpha,l \alpha'}}{(k,\omega)}|^2.
      \end{aligned}
\end{equation}
This component can be rewritten as $J_\beta^{(1)} = J_\beta^P+J_\beta^{Th-P}$ where
\begin{itemize}
\item  [-] $J_\beta^P$ is the second term of Eq. \ref{ec:JLorduno}, being a contribution purely induced by the pumping due to the time-dependent perturbation. 
\item [-] $J_\beta^{Th-P}$ includes the first and third terms of Eq. \ref{ec:JLorduno} and constitutes a mixed component that accounts for an interference process between the temperature gradient and the pumping. It can be expressed as
\end{itemize}
\begin{equation}
 \begin{aligned}
    J_\beta^{Th-P}&=\sum_{\beta'} \sum_k \int \frac{d\omega}{2\pi}\left(-\frac{d\eta_\beta}{d\omega}\right)\frac{\Delta T}{T_\beta}k\omega_0 \omega \Gamma_{\beta'}\left[\Gamma_\beta+\omega \frac{d\Gamma_\beta}{d\omega}\right]|\mathcal{G}_{f_{l \alpha,l \alpha'}}{(k,\omega)}|^2,\\
    &=\frac{\Delta T}{T_\beta}J_\beta^P+
    \frac{\Delta T}{T_\beta}\sum_{\beta'} \sum_k \int \frac{d\omega}{2\pi}\left(-\frac{d\eta_\beta}{d\omega}\right)\omega^2 k \omega_0\Gamma_{\beta'}\frac{d\Gamma_\beta}{d\omega}|\mathcal{G}_{f_{l \alpha,l \alpha'}}{(k,\omega)}|^2.
\end{aligned}
\end{equation}  
References\cite{arrPRB1,Brouwer,Moskalets2} show that from properties of the dynamical scattering matrix (Appendix IV), $J_\beta^p$ satisfies
\begin{equation*}
   \sum_{\beta=L,R}J_\beta^P=0
\end{equation*}
Therefore if $J_L<0$, $J_R>0$, this component is responsible for the phonon pumping against the temperature gradient. In other words, $J_\beta^p$ is at the helm of the cooling mechanism generated by the energy exchange processes $\hbar \omega  \Longleftrightarrow \hbar (\omega \pm \omega_0)$.

On the other hand, the second-order contribution $J_\beta^{(2)}$ can be expressed as 
\begin{equation}
 \begin{aligned}
    \label{ec:JLorddos}
       J_\beta^{(2)}\approx &\sum_{\beta'} \sum_k\int\frac{d\omega}{2 \pi}\left(\frac{d \eta_\beta}{d\omega}\right) k^2 \omega_0^2  \Gamma_{\beta'}(\omega) \Gamma_\beta(\omega) |\mathcal{G}_{f_{l \alpha,l \alpha'}}{(k,\omega)}|^2+\\
        &+\sum_{\beta'} \sum_k\int\frac{d\omega}{2 \pi}
        \left(-\frac{d \eta_\beta}{d\omega}\right) \frac{\Delta T }{T_\beta} k^2\omega\omega_0^2 \Gamma_{\beta'}(\omega)\frac{d\Gamma_\beta}{d\omega} |\mathcal{G}_{f_{l \alpha,l \alpha'}}{(k,\omega)}|^2+\\
        &+\sum_{\beta'} \sum_k\int\frac{d\omega}{2 \pi}
        \left(\frac{d \eta_\beta}{d\omega}\right) k^2\omega\omega_0^2 \Gamma_{\beta'}(\omega)\frac{d\Gamma_\beta}{d\omega} |\mathcal{G}_{f_{l \alpha,l \alpha'}}{(k,\omega)}|^2.
\end{aligned}
\end{equation}
This contribution, defined as $J^D_{L/R}$, contains all the terms depending on $\omega_0^2$ and is related to the dissipation of heat generated by the mechanical perturbation into the reservoirs. When $\omega_0 \rightarrow 0$, becomes negligible.

From the expansion given in Eq. \ref{Ec.jalphaaprox} we can have an interesting insight of the phenomenology behind the behaviors observed in Figs. \ref{fig:J_vs_deltaT_N24} and \ref{fig:CPvsDT}.
In the adiabatic limit the current presents a linear dependence with $\Delta T$ (if $\Delta T$ is small), in agreement with the numerical results of Fig. \ref{fig:J_vs_deltaT_N24} obtained from the full expression of J given in Eq.~\ref{fullJ}. 
On one hand Fig. \Ref{fig:J_vs_deltaT_N24} shows that the linear response occurs for local and non-local cases. On the other hand, the cooling regime occurs as long $\Delta T $ is smaller than a $\Delta T_{max}$, which depends on the set of parameters. In this regime $J^P$ flows from cold to hot reservoir, consequently this is the dominant contribution to the total current. Meanwhile $J_0$, $J^{th-p}$ and $J^D$ are all positive and responsible for the transport in opposite direction (from hot to cold). As long, $\Delta T \rightarrow \Delta T_{max}$ heat flow due to dissipation and thermal-pumping interference increases.

In Fig. \ref{fig:CPvsDT} we plot $CP$ versus  $\Delta T$ for the full current given in Eq. \ref{fullJ} for two values of $\omega_0$ far from spectrum cut-off frequency.  In order to understand the mechanism behind the $CP$ behavior, we analyze the different contributions in Eq. \ref{Ec.jalphaaprox}. To this end, we calculate $CP$ using the expansion of $J_L$ y $J_R$ (Eq. \ref{Ec.jalphaaprox}) up to order $\omega_0^2$.
Considering Eqs. \ref{ec:JLordcero}, \ref{ec:JLorduno} y \ref{ec:JLorddos}, $CP$ can be expressed as:
\begin{equation*}
CP^{(2)}(\Delta T)=\frac{\mathcal{A}^{(2)} + \mathcal{B}^{(2)}}{\mathcal{C}^{(2)}+ \mathcal{D}^{(2)}}=\frac{\mathfrak{a}^{(2)}+ \mathfrak{b}^{(2)} \Delta T}{\mathfrak{c}^{(2)}+ \mathfrak{d}^{(2)} \Delta T}
\end{equation*}
where $^{(2)}$ indicates coefficients depending on contributions up to second order. After some algebra we define the following coefficients $\mathcal{A}^{(2)}=J_{cold}^{P}+J_{cold}^{D}$, with $J_\beta^{D}$  containing the terms of Eq. \ref{ec:JLorddos} not depending on $\Delta T$,  $\mathcal{B}^{(2)}= J_{cold}^{0}+J_{cold}^{Th-P}+J_{cold}^{D}(\Delta T)$, where $J_{cold}^{D}(\Delta T)$ is the component of Eq. \ref{ec:JLorddos} linearly dependent on  $\Delta T$, $\mathcal{C}^{(2)}=J_{hot}^{D}+J_{cold}^{D}$  independent of $\Delta T$, and finally $\mathcal{D}^{(2)}=J_{cold}^{Th-P}+J_{hot}^{Th-P}+J_{cold}^{D}(\Delta T)+J_{hot}^{D}(\Delta T)$ including contributions that depend on $\Delta T$.

In the limit of small $\omega_0$, the dissipative terms in Eq. \ref{Ec.jalphaaprox} are negligible, therefore after some algebra $CP$ turns out to be 
\begin{equation*}
CP^{(1)}(\Delta T)=\mathcal{A}^{(1)}+\mathcal{B}^{(1)} =\mathfrak{a}^{(1)}+\mathfrak{b}^{(1)}\Delta T^{-1}
\end{equation*}
with $^{(1)}$ denoting contributions up to order $\omega_0$ and  $\mathcal{A}^{(1)}=\frac{J_{cold}^{Th-P}}{J_{cold}^{Th-P}+J_{hot}^{Th-P}}$,  $\mathcal{B}^{(1)}=\frac{J_{cold}^{0}+J_{cold}^{P}}{J_{cold}^{Th-P}+J_{hot}^{Th-P}}$. 
Consequently, in the adiabatic limit,  $CP$ obeys a power law dependence $\sim \Delta T^{-1}$.  This dependence is obtained if we neglect the term $\mathcal{C}^{(2)}$ in the denominator of $CP^{(2)}$.

Fig.\ref{fig:CPvsDT}(log-log) shows that when the frequency is small ($\omega_0/2\pi=0.05$),  $CP$ resembles the functional form given by $\sim CP^{(1)}$ (dotted line) in most of the interval $\Delta T< \Delta T_{max}$. Moreover, the curves corresponding to the local and non-local cases are practically coincident. This fact indicates that when the dissipative effects are negligible, the distinction between locality and non locality do not play a central role in the cooling performance of the system.


On the other hand, when the frequency increases ($\omega_0/2\pi=0.15$) there is a good agreement between the $CP$ estimation and the functional form $CP^{(2)}$. Moreover, the separation between the curves for local and non-local interactions becomes more pronounced and the range of $\Delta T$ where dissipation is negligible is remarkably reduced. Therefore, the effect of non locality on the cooling performance is enhanced when dissipation is stronger. In consequence, only when $\Delta T << \Delta T_{max} $, $CP^{(1)}$ results a good approximation.
It is interesting to note that for a given set of parameters, it is possible to obtain the same $CP$ value, but it is achieved for a larger $\Delta T$ when local interactions are considered. In other words, although the cooling performance is the same, the non locality reduces the temperature difference for an operative refrigeration regime.

\section{Conclusions}
We presented a microscopic model to study the heat transfer through a system formed by three chains in contact through a time-dependent mechanical coupling and also coupled to thermal reservoirs in both external segments. 
Using the Keldysh non equilibrium Green function formalism, we  calculated  the stationary heat current, finding the existence of different dynamical transport regimes. Depending on the characteristic parameters, we found that the system can act as a local thermal insulator, a heat engine and as a refrigerator, pumping energy from low to high temperatures.  Considering non-local interactions, these regimes displays qualitative changes. Regarding the cooling one, the non locality reduces the range of temperature gradients in which the system acts as a thermal refrigerator. From an adiabatic analysis we showed that non locality enhances the effect of dissipation, with the consequent reduction of the cooling performance of the system.

Although our results are obtained in the low frequency approximation, they shed  light  on  the  role  of dissipation  and  locality  on  the  cooling  performance  that  goes  beyond  the adiabatic limit. 
Usually, first principle studies regarding thermal transport in time-dependent one-dimensional systems are based on first-neighbors interacting models. Our results attempt to show that the choice of local interaction models must be carried out with a critical analysis to avoid distortions or underestimations of the dissipative effects. This can be a sensitive issue regarding the modeling of thermoelectric devices or low dimensional thermal devices.


\section*{Appendix I}
The Hamiltonian of Eq. \ref{res_normal} can be written in terms of normal modes for open boundary conditions
\begin{equation}
\label{mnR}
x_{i,\beta}= \sqrt{\frac{2}{N_{\beta}+1}} \sum_{k_{\beta}=0}^{N_\beta} \sin(u_{k_{\beta}} i) x_{k_{\beta}},
\end{equation}
and
\begin{equation}
\label{pnr}
p_{i,\beta}=\sqrt{ \frac{2}{N_{\beta}+1}} \sum_{k_{\beta}=0}^{N_\beta} \sin(u_{k_{\beta}} i) p_{k_{\beta}},
\end{equation}
with
\begin{equation}
\label{unr}
u_{k_{\beta}}= \frac{k_{\beta} \pi}{N_\beta +1},\hspace{1cm}  k_\beta=0\:,...\:,N_\beta.
\end{equation}
Then, the Hamiltonian of the reservoirs 
\begin{equation}
\label{res_normal}
H_{\beta}= \sum_{k_{\beta}=0}^{N_\beta} \frac{p_{k\beta}^2}{2 m_{k\beta}}+\frac{1}{2} K_{\beta}[1- \cos(u_{k_{\beta}}) ] x_{k_{\beta}}^2, \end{equation}
The coupling between the reservoir particles and the central chain can be expressed as
\begin{eqnarray}
 \label{eqnacoplam}
 \gamma_{k_{\beta},{l_{\alpha}}} =  \frac{K_{\beta c}}{m_{\alpha}}\sqrt{ \frac{2}{(N_{\beta} +1)}}\sin(u_{k_\beta}).
\end{eqnarray}


\section*{Appendix II}
The force matrix corresponding to our model is

\begin{equation}
F^{0}_{k,k'}(t) = \left\{ \begin{array}{cc} 
\frac{K_{I}}{m_I}( 2\delta_{k,k'}-\delta_{k',k\pm1}) &  1<k<N/4   \\ \\
\frac{K_{II}}{m_{II}}( 2\delta_{k,k'}-\delta_{k',k\pm1}) &  N/4+1<k<3N/4 \\ \\
\frac{K_{III}}{m_{III}}( 2\delta_{k,k'}-\delta_{k',k\pm1}) &  3N/4+1<k<N \\ \\
\left(\frac{K_{I}}{m_I}+\frac{K_{LC}}{m_L}\right)\delta_{k,k'}- K_{I} \delta_{k',k\pm1}&  k=1 \\ \\
\left(\frac{K_{III}}{m_{III}}+\frac{K_{RC}}{m_R}\right)\delta_{k,k'}- K_{III} \delta_{k',k\pm1} &  k=N \\ \\
\left(\frac{K_{I}}{m_I}+\frac{K_{1,int}^{0}}{m_{II}}\right)(\delta_{k,k'}- \delta_{k',k\pm1}) &  k=N/4 \\ \\
\left(\frac{K_{II}}{m_{II}}+\frac{K_{1,int}^{0}}{m_{II}}\right)(\delta_{k,k'}- \delta_{k',k\pm1}) &  k=N/4+1\\ \\
\left(\frac{K_{II}}{m_{II}}+\frac{K_{2,int}^{0}}{m_{III}}\right)(\delta_{k,k'}- \delta_{k',k\pm1}) &  k=3N/4 \\ \\
\left(\frac{K_{III}}{m_{III}}+\frac{K_{2,int}^{0}}{m_{III}}\right)(\delta_{k,k'}- \delta_{k',k\pm1}) &  k=3N/4+1
\end{array}\right. 
\end{equation}
\begin{equation}
F^{0(NL)}_{k,k'} (t)= \left\{ \begin{array}{cc} 
\frac{K^{I,\nu}}{m_{I}}( -\delta_{k,k'+2}-\delta_{k',k\pm2}) &  3<k<N/4+2 \\ \\
\frac{K^{{II},\nu}}{m_{II}}( -\delta_{k,k'+2}-\delta_{k',k\pm2}) &  N/4+3<k<3N/4-2 \\ \\
\frac{K^{{III},\nu}}{m_{III}}( -\delta_{k,k'+2}-\delta_{k',k\pm2}) &  3N/4+3<k<N-2 
\end{array}\right. 
\end{equation}
\begin{equation}
Fo^{0}_{k,k'}(t) = - K_{0} \delta_{k,k'} 
\end{equation}
\begin{equation}
F^{1}_{k,k'}(t) = \left\{ \begin{array}{cc} 
\frac{1}{2}\left(\frac{K_{I}}{m_I}+\frac{K_{1,int}^{1}}{m_{II}}\right) (\delta_{k,k'}- \delta_{k',k\pm1}) &  k=N/4 \\ \\
\frac{1}{2}\left(\frac{K_{II}}{m_{II}}+\frac{K_{1,int}^{1}}{m_{II}}\right)(\delta_{k,k'}- \delta_{k',k\pm1}) &  k=N/4+1 \\ \\
\frac{1}{2}\left(\frac{K_{II}}{m_{II}}+\frac{K_{2,int}^{1}}{m_{III}}\right)(\delta_{k,k'}- \delta_{k',k\pm1}) &  k=3N/4 \\ \\ 
\frac{1}{2}\left(\frac{K_{III}}{m_{III}}+\frac{K_{2,int}^{1}}{m_{III}}\right)(\delta_{k,k'}-\delta_{k',k\pm1}) &  k=3N/4+1
\end{array}\right. 
\end{equation}
with $K_{1,int}^{1}$ and $K_{2,int}^{1}$ are coupling constants between segments $I-II$ and $II-III$ respectively (Eq. \ref{ec.acoplamiento2}).


\section*{Appendix III}
The self-energy in Eq. (\ref{DysonEq}) can be expressed as
\begin{equation}
\Sigma^{R}_{k_\alpha,{k'}_{\alpha}}(t,t')=\sum_{\beta=L,R}\delta_{k'_{\alpha},{l'}_{\alpha}}\delta_{k_{\alpha},l_{\alpha}}\int_{-\infty}^{\infty}\frac{d\omega}{2\pi}
e^{-i \omega (t-t')} \times 
\int_{-\infty}^{\infty}\frac{d\omega'}{2\pi}\frac{\Gamma_{\beta}(\omega')}{\omega- \omega' + i \eta},
\end{equation}
with $\eta >0$. $\Gamma_{\beta}(\omega)$ is the spectral density of the reservoir $\beta$ 
\begin{align}
\label{gamma}
  \Gamma_{\beta}(\omega)&=\lim_{N_{\beta} \rightarrow \infty}\frac{2\pi K_{\beta c}^2}{m_{\beta} m_{\alpha}(N_{\beta}+1)} 
  \sum_{k_{\beta}=0}^{N_{\beta}} \sin^2\left(u_{k_\beta}\right)\frac{1}{\omega_{k_\beta}}
  \left[ \delta\left( \omega-\omega_{k_\beta}\right) +\delta\left( \omega+\omega_{k\beta}\right) \right],\nonumber\\
  &=\sgn\left(\omega\right)\left(\frac{K_{\beta c}}{m_{\alpha}}\right) \Theta\left(1-\left(\frac{K_{\beta c }-m_{\beta}\omega^2}{K_{\beta c}}\right)^2\right)
  \sqrt{1-\left(\frac{K_{\beta c}-m_{\beta}\omega^2}{K_{\beta c}}\right)^2}.
\end{align}

\section*{Appendix IV}
Using the relationship between the frozen scattering matrix and the frozen Green function \cite{arrPRB} 
\begin{equation}
    \mathcal{S}^f_{\beta \beta'}(k,\omega)=\delta_{\beta\beta'}\delta_{k,0} -i \sqrt{\Gamma_\beta(\omega) \Gamma_{\beta'}(\omega)} \mathcal{G}_{f_{l\alpha,l \alpha'}}{(k,\omega)}.
\end{equation}
$J_\beta^P$ can be written as
\begin{equation}
    J_\beta^P=\frac{1}{\tau}\int_0^\tau dt \int \frac{d\omega}{2\pi}\omega \left(-\frac{\partial \eta_\beta(\omega)}{\partial \omega}\right) \Im \left[{\mathcal{S}^f}(t, \omega)\partial_t{\mathcal{S}^f}^\dagger(t,\omega)\right].
\end{equation}
Applying the Birman-Krein relation 
 $d\ln([det\hat{S}])=Tr[\hat{S}d\hat{S^\dagger}]$($det(\hat{Z})$ and $Tr(\hat{Z})$ denote respectively the determinant and the trace of a matrix $\hat{Z}$) applied to the unitary frozen matrix~\cite{arrPRB} the following equality is fulfilled 
\begin{equation*}
   \sum_{\beta=L,R}J_\beta^P=0.
\end{equation*}

\section*{Acknowledgment}

M.F.C., N.B., A.S are supported by PIO Conicet. We thank TUPAC cluster of the Computational Simulation Center-CONICET.


\bibliographystyle{plain}

\end{document}